\documentclass[10pt,conference]{IEEEtran}
\usepackage{cite}
\usepackage{amsmath,amssymb,amsfonts}
\usepackage{algorithmic}
\usepackage{graphicx}
\usepackage{textcomp}
\usepackage[hyphens]{url}
\usepackage{fancyhdr}
\usepackage{hyperref}
\usepackage{listings}
\usepackage{subcaption}
\usepackage[table]{xcolor}
\usepackage{array}
\usepackage{xcolor}
\usepackage{multicol}
\usepackage{tabularx}
\usepackage[font=small]{caption} % Set the font size for captions to 9pt
\definecolor{codegray}{rgb}{0.4,0.4,0.4}   % Gray used for line numbers
\definecolor{codeblue}{rgb}{0,0,0.8}       % Blue used for keywords
\definecolor{codegreen}{rgb}{0,0.5,0}      % Green used for comments
\definecolor{codemaroon}{rgb}{0.5,0,0}     % Maroon used for strings

\lstdefinestyle{mystyle}{
    backgroundcolor=\color{white},          % White background for the code block
    commentstyle=\color{codegreen},         % Green style for comments within the code
    keywordstyle=\color{codeblue}\bfseries, % Blue and bold for keywords within the code
    numberstyle=\footnotesize\color{codegray},      % Footnotesize, gray for the line numbers
    stringstyle=\color{codemaroon},         % Maroon for string literals in the code
    basicstyle=\normalfont\footnotesize\ttfamily,  % Footnotesize and monospace font for the code
    breakatwhitespace=false,                % Line breaks happen at whitespace
    breaklines=true,                        % Enable line wrapping
    captionpos=b,                           % Caption is positioned below the code block
    keepspaces=true,                        % Preserve spaces in the code, useful for alignment
    numbers=left,                           % Line numbers on the left side of the code block
    numbersep=10pt,                         % Increased space between line numbers and the code
    showspaces=false,                       % Do not show space characters
    showstringspaces=false,                 % Do not show spaces within strings
    showtabs=false,                         % Do not show tab characters
    tabsize=2,                              % Tab character is equal to two spaces
    frame=single,                           % Single frame around the code block
    framexleftmargin=5pt,                   % Increased margin on the left side of the frame
    framexrightmargin=5pt,                  % Increased margin on the right side of the frame
    xleftmargin=10pt,                       % Increased margin on the left side outside the frame
    xrightmargin=10pt,                      % Increased margin on the right side outside the frame
    framesep=2pt,                           % Separation between the frame and the code
    framerule=0.5pt,                        % Width of the frame line
    belowcaptionskip=0pt, % Set margin between frame and caption to 0
    abovecaptionskip=0pt, % Set margin between frame and caption to 0
    moredelim=[is][\bfseries\color{codegray}]{\#}{\#} % Ensures bold gray for the line number
}

\usepackage{tikz}
\usepackage{xprintlen}

\newcommand*\circled[1]{\tikz[baseline=(char.base)]{
            \node[shape=circle,fill=black,text=white,inner sep=1pt] (char) {#1};}}

\makeatletter
\AtBeginDocument{\let\c@lstlisting\c@figure}
\def\lst@PlaceNumber{\llap{{\normalfont \normalcolor \lst@numberstyle \thelstnumber :}\kern \lst@numbersep}}
\makeatother

\DeclareCaptionFormat{listing}{\textbf{Fig.~\thefigure:} #3}
\graphicspath{{./figs/}}

\newcommand{\hpcayear}{2024}

\newcommand{\hpcasubmissionnumber}{NaN}
\title{Control Flow Management in Modern GPUs}
\def\hpcacameraready{} % Uncomment to build camera-ready version

\newcommand\hpcaauthors{Mojtaba Abaie Shoushtary$\dagger$ and Jordi Tubella Murgadas$\dagger$ and Antonio Gonzalez$\dagger$}
\newcommand\hpcaaffiliation{Universitat Politècnica de Catalunya (UPC)$\dagger$}
\newcommand\hpcaemail{mojtaba.abaie@upc.edu, jordi.tubella@upc.edu, antonio@ac.upc.edu}

\author{
  \ifdefined\hpcacameraready
    \IEEEauthorblockN{\hpcaauthors{}}
      \IEEEauthorblockA{
        \hpcaaffiliation{} \\
        \hpcaemail{}
      }
  \else
    \IEEEauthorblockN{\normalsize{HPCA \hpcayear{} Submission
      \textbf{\#\hpcasubmissionnumber{}}} \\
      \IEEEauthorblockA{
        Confidential Draft \\
        Do NOT Distribute!!
      }
    }
  \fi 
}

\fancypagestyle{camerareadyfirstpage}{%
  \fancyhead{}
  
  \fancyhead[C]{
    \ifdefined\aeopen
    \parbox[][12mm][t]{13.5cm}{\hpcayear{} IEEE International Symposium on High-Performance Computer Architecture (HPCA)}    
    \else
      \ifdefined\aereviewed
      \parbox[][12mm][t]{13.5cm}{\hpcayear{} IEEE International Symposium on High-Performance Computer Architecture (HPCA)}
      \else
      \ifdefined\aereproduced
      \parbox[][12mm][t]{13.5cm}{\hpcayear{} IEEE International Symposium on High-Performance Computer Architecture (HPCA)}
      \else
      \parbox[][0mm][t]{13.5cm}{\hpcayear{} IEEE International Symposium on High-Performance Computer Architecture (HPCA)}
    \fi 
    \fi 
    \fi 
    \ifdefined\aeopen 
      \includegraphics[width=12mm,height=12mm]{ae-badges/open-research-objects.pdf}
    \fi 
    \ifdefined\aereviewed
      \includegraphics[width=12mm,height=12mm]{ae-badges/research-objects-reviewed.pdf}
    \fi 
    \ifdefined\aereproduced
      \includegraphics[width=12mm,height=12mm]{ae-badges/results-reproduced.pdf}
    \fi
  }
  \fancyfoot[C]{}
}
\begin{document}
\maketitle

%Enables the camera ready header and footer
% \ifdefined\hpcacameraready 
%   \thispagestyle{camerareadyfirstpage}
%   \pagestyle{empty}
% \else
  \thispagestyle{plain}
  \pagestyle{plain}
% \fi

\newcommand{\hpcaheight}{0mm}
\ifdefined\eaopen
\renewcommand{\hpcaheight}{12mm}
\fi

%%%%%% -- PAPER CONTENT STARTS-- %%%%%%%%

\begin{abstract}

In GPUs, the control flow management mechanism determines which threads in a warp are active at any point in time. This mechanism monitors the control flow of scalar threads within a warp to optimize thread scheduling and plays a critical role in the utilization of execution resources. The control flow management mechanism can be controlled or assisted by software through instructions. However, GPU vendors do not disclose details about their compiler, ISA, or hardware implementations. This lack of transparency makes it challenging for researchers to understand how the control flow management mechanism functions, is implemented, or is assisted by software, which is crucial when it significantly affects their research. It is also problematic for performance modeling of GPUs, as one can only rely on traces from real hardware for control flow and cannot model or modify the functionality of the mechanism altering it.

This paper addresses this issue by defining a plausible semantic for control flow instructions in the Turing native ISA based on insights gleaned from experimental data using various benchmarks. Based on these definitions, we propose a low-cost mechanism for efficient control flow management named Hanoi. Hanoi ensures correctness and generates a control flow that is very close to real hardware. Our evaluation shows that the discrepancy between the control flow trace of real hardware and our mechanism is only 1.03\% on average. Furthermore, when comparing the Instructions Per Cycle (IPC) of GPUs employing Hanoi with the native control flow management of actual hardware, the average difference is just 0.19\%.

\end{abstract}
\section{Introduction}

Graphics Processing Units (GPUs) employ a Single Instruction Multiple Threads (SIMT) \cite{Tesla} architecture that executes multiple threads simultaneously on Single Instruction Multiple Data (SIMD) processing units. SIMD lanes in this architecture perform identical operations on different operands from distinct threads. Thread scheduling significantly impacts SIMD utilization and the overall performance of such architecture, as only the scheduled threads can utilize the SIMD lanes.

GPUs leverage two mechanisms for scheduling threads for execution: a) threads are grouped into sets named warps \cite{Tesla,CudaGuide} and every cycle a warp is selected for execution \cite{CCWS}, and b)  determining which threads in a warp are active at any point in time. We call the latter mechanism control flow management primarily because it determines the control flow of each warp and is highly influenced by the control flow of individual threads. The control flow management mechanism monitors the control flow of threads within a warp and co-schedules threads executing the same instruction. This mechanism can be controlled or assisted by software through instructions in the Instruction Set Architecture (ISA) to perform optimal and efficient thread scheduling.  

Designing an efficient control flow management mechanism is critical and challenging for modern GPUs, primarily because modern GPUs support a rich set of control-flow instructions. For instance, the NVIDIA Turing architecture \cite{Turingwhitepaper} includes 20 control-flow instructions in its native ISA, called SASS \cite{TuringSASS}. These control flow instructions may diverge threads to different paths, which makes it impossible to schedule them together as they execute different instructions. The divergence of threads reduces SIMD utilization and performance; therefore, the control flow management mechanisms leverage the runtime information of individual threads' control flow and software information to reunite threads to gain higher efficiency.

Researchers have developed numerous software and hardware mechanisms to manage control flow \cite{DWF,TBL_Compaction,Large_Warp,Multi-Path,subwarp_interleaving,MIMD_Sync,Dual_Path,DWS,Tesla,RPU,Thread_Frontiers} that were designed based on publicly available information and evaluated with open-source tools. They have extensively used LLVM \cite{LLVM} for compiler implementations and GPGPU-Sim 3.x \cite{GPGPU-Sim_3} as a performance model for evaluations. Both GPGPU-Sim 3.x and LLVM utilize the well-documented Parallel Thread Execution (PTX) ISA \cite{PTX} as the interface between hardware and software. Researchers have used this strategy for years to address the challenges of the opaque compiler, ISA, and hardware imposed by leading GPU vendors such as NVIDIA.

One of the major problems of this approach is that it relies on PTX as the ISA since it is well documented. Therefore, it implies modeling and optimizing microarchitectures running this ISA. However, PTX is different from what GPUs actually run, so these models may significantly deviate from real hardware. This deviation is significant when it comes to control flow management. This is because NVIDIA GPUs run a native ISA called SASS, not PTX, and the translation from PTX to SASS is not near one-to-one as it was in the early generations of NVIDIA GPUs \cite{Tesla}. Considering only control-flow instructions, PTX has 5 instructions while SASS ISA has 20 in Turing.
Furthermore, a PTX code goes through static optimizations before generating the final SASS, which may change the final control flow. When one designs a control flow management mechanism based on PTX, there are constraints that are ignored. These constraints are posed by instructions in the SASS that need to be supported, but they do not exist in PTX. Furthermore, to study the state-of-the-art implementation of modern workloads such as deep learning \cite{mlpef_inference,mlperf_training,Deepbench} or graph analytics \cite{GraphBIG}, we have no choice but to rely on SASS ISA since these applications use fine-tuned libraries such as cuDNN \cite{cuDNN}, cuBLAS \cite{cuBLAS}, etc. These libraries are highly optimized and provided by NVIDIA. However, as their source code is not available, one can only study them by profiling or collecting SASS traces when they run on real hardware.

Researchers have acknowledged the limitations of simulating PTX and have developed trace-driven simulators such as Accel-Sim \cite{Accel-Sim}. Accel-Sim simulates traces of SASS instructions running on modern GPU architectures like Volta \cite{Volta_Whitepaper} and Turing. Like most trace-driven simulators, only the performance of the hardware components is modeled, not their functionality. This is basically because the functionality of the instructions and hardware components is not disclosed either, and discovering the underlying hardware mechanisms requires extra effort and is not always feasible. As a result, simulators such as Accel-Sim rely only on the control flow traces that real hardware generates for the control flow management mechanism.

However, when researchers develop a mechanism that changes the functionality of hardware components, simply simulating performance or relying on hardware traces does not work. This is particularly true when working on new control flow management mechanisms, where a new scheme's functionality alters the control flow and has side effects that affect other components like issue schedulers and dependence checking mechanisms. Therefore, the functionality of the control-flow management mechanism needs to be modeled to study its effect and evaluate alternative designs of it.

For modeling the functionality of the control-flow management mechanism, one first needs to know the semantics of control-flow instructions in the ISA and then the microarchitecture details. None of these details are publicly available. Even related works dissecting the architecture and ISA of modern GPUs, such as  Volta \cite{Dissecting_Volta} and Turing \cite{dissecting_Turing}, have not addressed the control flow management mechanism. 

In this work, we bridge this gap by defining the semantics of control-flow instructions in the Turing ISA based on experimental data we collected by studying the binary and traces of various applications. This approach allows us to design a novel control flow management mechanism named Hanoi that supports these control-flow instructions. We demonstrate that the control flow generated by Hanoi produces the correct output for all benchmarks and closely mirrors real hardware. Specifically, comparing the control flow traces of real hardware and Hanoi shows a 1.03\% discrepancy on average, which leads to a minor 0.19\% change in performance.

To the best of our knowledge, Hanoi is the first control flow management mechanism that is designed to support all control-flow instructions in the Turing ISA that appear in a diverse set of well-known benchmarks. Hanoi is lightweight in terms of hardware cost and is proven to be highly similar to the actual hardware mechanism. Other proposed schemes \cite{DWF,TBL_Compaction,Multi-Path,Dual_Path,MIMD_Sync,DWS} are designed for PTX ISA, or their cost/benefits do not justify their use in a real product \cite{subwarp_interleaving}. Furthermore, this is the first attempt to describe plausible semantics for control-flow instructions in Turing ISA that appear in common benchmarks. The semantics of a few control-flow instructions in Turing that are mentioned in the literature \cite{subwarp_interleaving} is incomplete as only 3 out of 20 instructions are defined. Only these instructions cannot handle all the scenarios we encountered in the benchmarks.

To sum up, the major contributions of this paper are the following:
\begin{itemize}
    \item We define the semantics of control-flow instructions in the Turing ISA.
    \item We design a novel control flow management mechanism for Turing named Hanoi.
    \item We compare Hanoi and the control flow management mechanism implemented in Turing actual hardware and prove their extremely high similarity. Their control flow trace differs only by 1.03\% on average, leading to a minor 0.19\% IPC change.
\end{itemize}

\section{Pre-Volta Control Flow Management}
\begin{figure}[t]
    \centering
    \begin{subfigure}[b]{0.44\columnwidth}
        \centering
        \begin{lstlisting}[language=C, aboveskip=0pt, belowskip=0pt, , escapeinside={(*@}{@*)}]
A;
(*@\label{line:diverge}@*)if (tid % 4 < 2) {
    B;
    C;
} else {
    D;
    E;
}
(*@\label{line:converge}@*)F;\end{lstlisting}
        \caption{}
        \label{fig:pre_volta_divergent_code}
    \end{subfigure}
    \hfill
    \begin{subfigure}[b]{0.203\columnwidth}
        \centering
        \includegraphics[width=\linewidth]{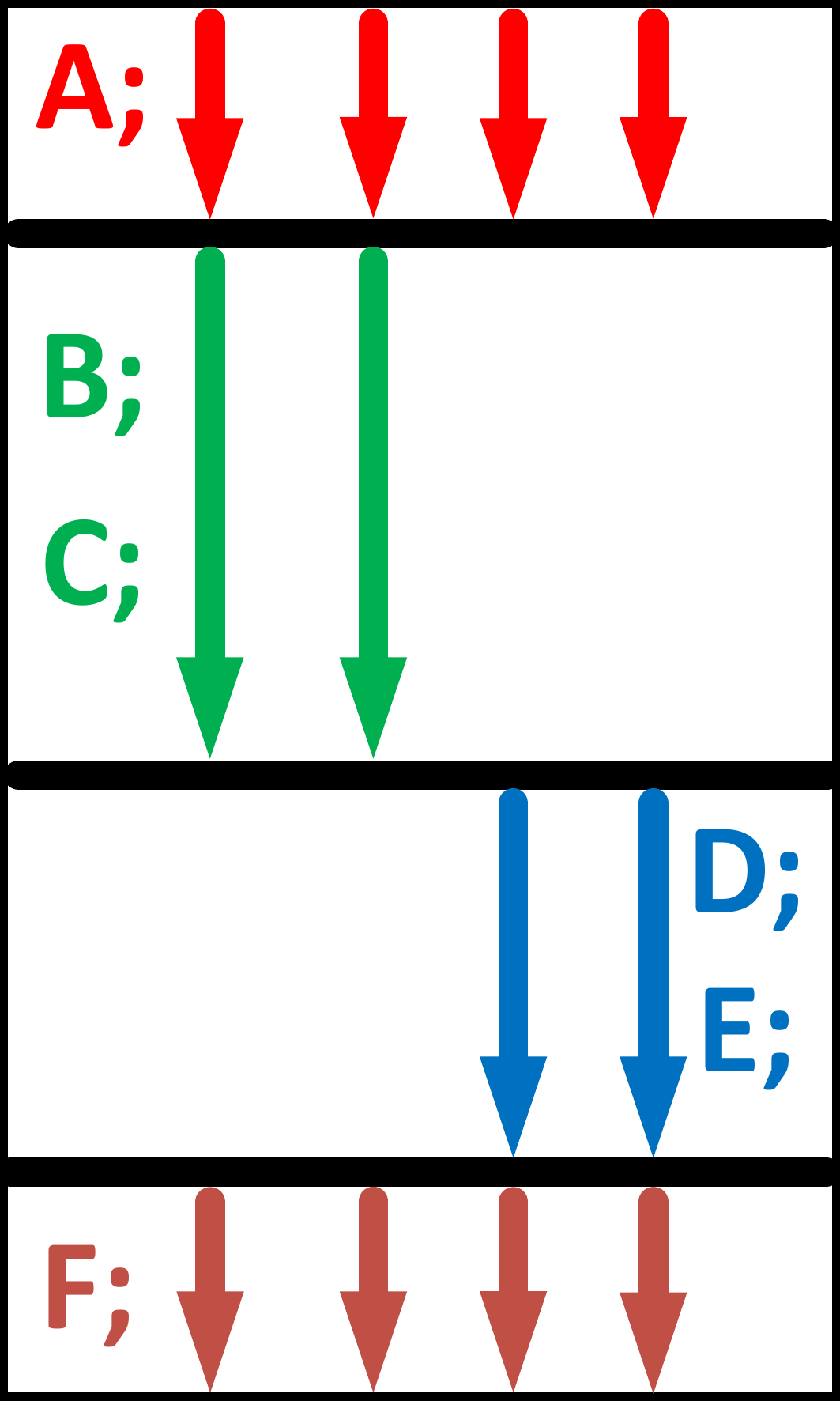}
        \caption{}
        \label{fig:pre_volta_execution_model}
    \end{subfigure}   
    \hfill
    \begin{subfigure}[b]{0.305\columnwidth}
        \centering
        \includegraphics[width=\linewidth]{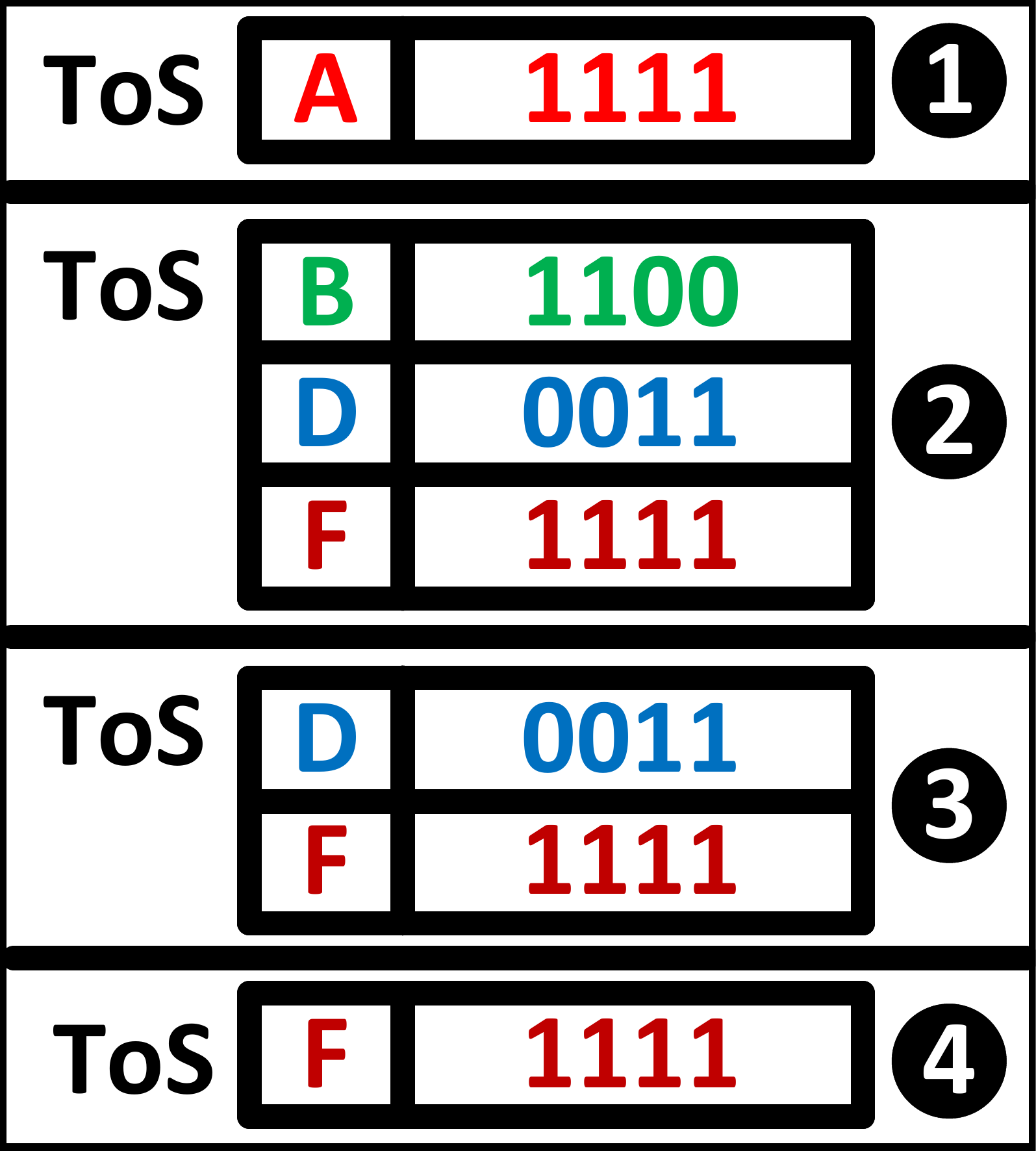}
        \caption{}
        \label{fig:pre_volta_cfmg_implementation}
    \end{subfigure}
    \caption{Control flow management for a pre-Volta GPU model with 4 threads in a warp: (a) Code sample with branch divergence \cite{independent_tsch}, (b) Divergent threads execution model \cite{CudaGuide,independent_tsch}, and (c) Plausible control flow management implementation \cite{RPU,DWF,GPGPU_SIM_Manual}}
    \label{fig:pre_volta_cfmg}
\end{figure}

Leading GPU vendors such as NVIDIA have disclosed some aspects of the control flow management mechanism to programmers through their documented execution model. The execution model provides the necessary assumptions for writing correct and deadlock-free programs in programming languages such as CUDA or PTX. Moreover, it offers valuable insights for researchers to infer a plausible implementation of control flow management, albeit the implementation details have never been disclosed.

Figure \ref{fig:pre_volta_cfmg} depicts a plausible control flow management for divergent threads for pre-Volta GPUs derived from the execution model. For simplicity, we assume a warp contains only four threads throughout this paper for illustration purposes. All threads in a warp start executing the source code shown in Figure \ref{fig:pre_volta_divergent_code} in lockstep. Once the threads execute line \ref{line:diverge}, branch divergence causes that the first half of the threads in the warp  follow the taken path while the remaining threads follow the not-taken.

In this case, the execution model specifications that NVIDIA provides \cite{CudaGuide,independent_tsch,Tesla} describe an execution as shown in Figure \ref{fig:pre_volta_execution_model}. This figure demonstrates a serial execution of the taken and not-taken paths, and posterior reconvergence at line \ref{line:converge}. The line \ref{line:converge} is a point called Immediate Post Dominator (IPDom) \cite{compiler_book}, which is the nearest point in the program where two diverging paths are guaranteed to converge again. Although the execution model does not specify which divergent path has priority, in this example the taken path is assumed to have higher priority than the not-taken one.

A plausible mechanism to achieve this behavior involves using a structure called SIMT-Stack \cite{RPU,DWF}. This mechanism utilizes the SIMT-Stack to track which threads are in which paths and enforce reconvergence at IPDom points. Each stack entry stores the next instruction's PC and the active mask of the threads executing that instruction. Threads with their corresponding bit set in the active mask execute the next instruction while the remaining threads stay idle.

Figure \ref{fig:pre_volta_cfmg_implementation} shows how the stack is updated when executing this sample code. Initially, the stack has only one entry showing all threads executing instruction A together (\circled{1}). When a branch divergence occurs, the entry on top of the stack will be popped, and three new entries will be pushed into the stack (\circled{2}). The first entry contains the IPDom's PC and the complete active mask, as all threads must continue from IPDom after reconvergence. Then, two entries are pushed, one for threads following the taken path and another for the ones following not-taken path.

After the taken path finishes, its entry will be popped, and execution continues with the not-taken path (\circled{3}). When threads finish the execution of the not-taken path, its entry is popped, and now the pointer of the top of the stack points to the entry with IPDom PC and full active mask (\circled{4}). This is how reconvergence at IPDom is enforced.

Although the SIMT-Stack is a derivation from the execution model, the model does not expose any deeper details about the control flow management mechanism, such as whether the stack is implemented purely in hardware or managed by software. Moreover, the model remains silent on the subtleties of the SIMT-Stack that could significantly influence performance, including the variety of path selection policies — whether to prioritize the taken path, the not-taken path, or to use a heuristic for path selection in hardware or software.

\begin{figure}[t]
    \centering
    \includegraphics[width=\columnwidth]{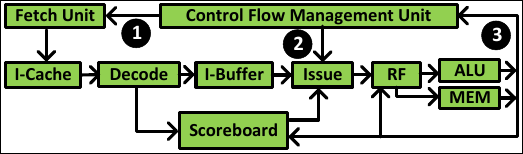}
    \caption{GPU's SIMT core microarchitecture \cite{GPGPU_SIM_Manual}}
    \label{fig:SIMT_Core}
\end{figure}

Researchers employed micro-benchmarking to uncover subtleties such as the path selection policy \cite{Dem_GPU_Micro}. They also widely adopted a SIMT stack that is entirely implemented in hardware and receives a minor assistance from software to identify the location of IPDom points for each branch \cite{GPGPU-Sim_3, GPGPU_SIM_Manual, DWF}. The researchers also made assumptions about how this mechanism is integrated into the GPU's core microarchtiecture.

Figure \ref{fig:SIMT_Core} illustrates a common assumption among researchers regarding the integration of this mechanism into the GPU's SIMT cores through dedicated units per warp, referred to as Control Flow Management Units (CFUs) \cite{GPGPU_SIM_Manual}. Each CFU encompasses a stack and some control logic. The CFU provides the PC and active mask for the next instruction. The fetch unit uses this PC (\circled{1}) to request instructions from memory. Once instructions are fetched and decoded, they will be stored in the Instruction Buffer (I-Buffer) slots. Each warp has a dedicated set of slots in the I-Buffer. In this SIMT core architecture, a warp issues instructions in the program order when it is selected to do so. The issue logic selects a particular warp whose next instruction does not have data/structural hazard. For data hazards, it utilizes a private scoreboard per warp for dependence checking.

When an instruction is issued, the active mask indicates which threads are executing this instruction. The CFU provides this mask (\circled{2}), which is used by the components in the following pipeline stages, for instance, to mask updates of registers or memory. A control-flow instruction may change the flow of control or the internal state of the CFU; therefore, the CFU needs to be updated after executing these instructions (\circled{3}). 

Leveraging the SIMT stack for control flow management has significantly streamlined the SIMT core architecture. This simplicity is primarily a consequence of the fact that the stack-based implementation does not support interleaved execution of different paths, eliminating the need for per-path I-Buffer slots or scoreboards as proposed elsewhere in the literature \cite{Multi-Path}. However, SIMT-Stack put constraints on thread scheduling which may cause deadlock in some scenarios and must be considered by programmers.

\section{SIMT-Induced Deadlocks in Pre-Volta}
\label{sec:pre_volta_deadlock}
\begin{figure}[t]
    \centering
        \begin{lstlisting}[language=C, aboveskip=0pt, belowskip=0pt, , escapeinside={(*@}{@*)}]
*mutex = 0;
while(!atomicCAS(mutex,0,1));        (*@\tikz[remember picture] \coordinate (code-start);@*)
(*@\label{line:ipdom}@*)// Critical Section
(*@\label{line:unlock}@*)atomicExch(mutex,0);                 (*@\tikz[remember picture] \coordinate (code-end);@*)\end{lstlisting}
\begin{tikzpicture}[remember picture,overlay]
    % Arrow annotation
    \draw[->, >=stealth, line width=0.5mm] (code-start) -- (code-end) node[left,midway] {path a};
    
    % Egg shape annotation
        \draw[->, >=stealth, line width=0.5mm] (code-start) arc (-180:180:0.45) node[midway, right] {path b};
\end{tikzpicture}
    \caption{Spinlock Cuda Implemetation}
    \label{fig:pre_volta_deadlock}
\end{figure}
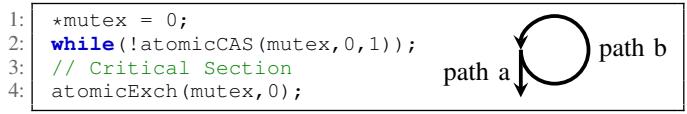

The pre-Volta execution model imposes three constraints on thread scheduling when thread divergence occurs. These constraints arise from the implementation of the control flow management mechanism. Programmers must carefully consider these constraints to ensure program correctness and avoid deadlocks. The constraints are as follows:
\begin{enumerate}
    \item Divergent paths are executed serially, one after another.
    \item Threads in each path are executed in a lockstep fashion.
    \item Reconvergence is enforced at the immediate post-dominator (IPDom) point.
\end{enumerate}

In the literature, deadlocks caused by SIMT implementation constraints are referred to as SIMT-induced deadlocks \cite{MIMD_Sync, Dem_GPU_Micro,correct_SIMT,GPU_verify,GKLEE}. In pre-Volta GPUs, the control flow management mechanism is the primary cause of most SIMT-induced deadlocks. For example, a spinlock implementation in CUDA, as shown in Figure \ref{fig:pre_volta_deadlock}, results in deadlock due to the control flow management mechanism \cite{MIMD_Sync}.

In this example, when a thread acquires the lock, it exits the loop, executes the critical section, and releases the lock for another thread to enter. Threads competing to acquire the lock diverge into two paths: one exiting the loop and proceeding to the critical section (path a), and the other returning to the beginning of the loop (path b). The thread acquiring the lock takes path a, while the remaining threads in the warp take path b, where they remain trapped in the loop until the lock is released.

This scenario results in a deadlock in pre-Volta GPUs due to constraints 1 and 3. According to constraint 1, one path must be executed before the other. If path b is given priority, a deadlock occurs because path b waits indefinitely for the lock to be released by path a (at line \ref{line:unlock}), which never happens. Conversely, if path a is prioritized, deadlock persists due to constraint 3, which requires reconvergence at the IPDom point (line \ref{line:ipdom}, right after the loop exit) before lock release (line \ref{line:unlock}). However, reconvergence cannot occur because threads waiting on the lock are stuck in the loop indefinitely since the thread that acquired the lock is block in the reconvergence point waiting for the other threads and never reaches the point where the lock is released.

This deadlock could not happen if the control flow management constraints were different. To address this issue, NVIDIA removed these constraints in the post-Volta execution model specifications. This indicates that post-Volta GPUs employ a substantially different control flow management mechanism, which has not been disclosed.
\section{Post-Volta Control Flow Management}
\begin{figure}[t]
    \centering
    \begin{subfigure}[b]{0.45\columnwidth}
        \centering
        \begin{lstlisting}[language=C, aboveskip=0pt, belowskip=0pt, , escapeinside={(*@}{@*)}]
A;
if (tid % 4 < 2) {
    B;
    C;
} else {
    D;
    E;
}
(*@\label{line:post_volta_convergence}@*)F;
(*@\label{line:syncwarp}@*)__syncwarp();
G;
\end{lstlisting}
        \caption{}
        \label{fig:post_volta_divergent_code}
    \end{subfigure}
    \hfill
    \begin{subfigure}[b]{0.45\columnwidth}
        \centering
        \includegraphics[width=\linewidth]{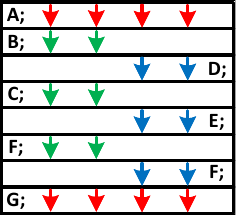}
        \caption{}
        \label{fig:post_volta_execution_model}
    \end{subfigure}
    \caption{Control flow management for a post-Volta GPU model with 4 threads in a warp: (a) Code sample with branch divergence \cite{independent_tsch}, (b) Plausible execution on post-Volta GPUs \cite{independent_tsch}}
    \label{fig:post_volta_cfmg}
\end{figure}

In the pre-Volta execution model, programmers were exposed to underlying control flow management constraints and had to intervene to avoid SIMT-induced deadlocks, making the process tedious. The introduction of Volta minimized the need for such intervention and simplified programming. The post-Volta execution model eliminated all three pre-Volta constraints and introduced only one new constraint. Therefore, programmers must now ensure that this single constraint does not cause cyclic dependencies between the execution of threads.

In the post-Volta execution model, divergent paths are not necessarily serialized, and threads of a warp do not have to be executed in lockstep manner. Furthermore, reconvergence at the IPDom point is no longer strictly enforced. Instead, a new scheduling mechanism called independent thread scheduling \cite{independent_tsch, CudaGuide, Volta_Whitepaper} is introduced, allowing any threads of a warp with the same PC to be scheduled together at any given cycle. This new scheduling method permits the interleaved execution of different paths and allows threads of a warp to diverge or converge at any instruction, not just at control-flow instructions or IPDom points. As a result, reconvergence can now occur earlier or later relative to the IPDom point in the execution flow or completely be ignored for some branches.

The sole constraint of this model is the enforcement of intra-warp synchronization at new instructions such as \_\_syncwarp that are introduced for this purpose \cite{CudaGuide,PTX,independent_tsch}. Consequently, programmers must take responsibility for synchronizing threads within a warp as needed to maintain correctness.

Figure \ref{fig:post_volta_cfmg} shows a sample source code similar to Figure \ref{fig:pre_volta_cfmg}; the main difference is the addition of a \_\_syncwarp() at line \ref{line:syncwarp} to synchronize all threads within a warp. Figure \ref{fig:post_volta_execution_model} illustrates a plausible execution of this source code on a post-Volta GPU. In this sample execution, threads in divergent paths are interleaved, and reconvergence at the IPDom point (line \ref{line:post_volta_convergence}) is ignored. However, all threads in the warp are synchronized after the execution of \_\_syncwarp. This example illustrates how programmers can enforce intra-warp synchronization.

This execution model simplifies programming but obscures almost all details of the underlying control flow management mechanism. For instance, it is clear that the control flow management mechanism in Turing is a hardware/software design since many control-flow instructions in the native ISA can assist it. However, the semantics of these instructions and how the mechanism leverages them are not disclosed. Furthermore, the detailed policies the control-flow management mechanism uses are not disclosed. For instance, while theoretically, threads can be scheduled independently in a post-Volta GPU, a more cost-effective scheduling policy might opt to schedule all threads in each path together and switch between paths occasionally to gain efficiency. This policy is just one of many plausible scheduling policies.

Since so many details are hidden from programmers in the post-Volta execution model, researchers face difficulties in inferring the underlying control flow management. This is due to the numerous plausible design decisions. Essentially, any hardware/software control flow management mechanism that does not result in pre-Volta SIMT-induced deadlocks and respects all intra-warp synchronizations is a plausible design. In this work, we aim to uncover the underlying control flow management mechanism by studying the binary and traces of a post-Volta GPU at its native ISA, SASS.

\section{Turing Control Flow Instructions}

We analyzed the native ISA of a Turing GPU to understand the underlying control flow management mechanism. The Turing ISA supports predication and contains 20 control flow instructions, summarized in Table \ref{tab:turing_instructions}. NVIDIA documents briefly mention these instructions but remain discreet about their functionality and semantics. We studied the binary and traces of various benchmarks to decipher these instructions. Only instructions highlighted in green appeared in our set of benchmarks. This section explains our findings and the rationale behind defining each instruction's semantics.

\begin{table}[t]
\centering
\begin{tabular}{|c|c|c|c|c|}
\hline
\cellcolor{green!30}BMOV & \cellcolor{green!30}BRA & \cellcolor{green!30}BREAK & BPT & BRX \\
\hline
\cellcolor{green!30}BSSY & \cellcolor{green!30}CALL & \cellcolor{green!30}EXIT & BRXU &  JMP  \\
\hline
\cellcolor{green!30}BSYNC & \cellcolor{green!30}WARPSYNC & RPCMOV & KILL & NANOSLEEP \\
\hline
\cellcolor{green!30}RET & \cellcolor{green!30}YIELD & RTT & JMX & JMXU \\
\hline
\end{tabular}
\caption{Control Flow Instructions in Turing's Native ISA \cite{TuringSASS}}
\label{tab:turing_instructions}
\end{table}

\subsection{Predicated Control Flow Instructions}
Our study shows that control flow instructions in Turing may be guarded by up to two predicate registers, which can be negated before use. These predicate registers are 32-bit registers for a warp with 32 threads, 1 bit per thread. The first predicate register is preceded by an \texttt{@} symbol and appears before the instruction, while the second predicate register is always the first operand. To negate a predicate, a \texttt{!} symbol must precede its register name. If an instruction has two predicates, a boolean \texttt{AND} operation is performed on them before predication. For example, \texttt{@P0 INST !P1, R0} indicates that the instruction will only be executed if \texttt{P0} is \texttt{true} and \texttt{P1} is \texttt{false}.

\subsection{EXIT Instruction}
The \texttt{EXIT} instruction terminates the thread’s execution. This instruction can have up to one predicate and has no operands. Threads that are masked by the predicate continue execution from the subsequent instruction, while the other threads are terminated.

\subsection{BRA Instruction}
The \texttt{BRA} instruction jumps to a target address conditionally or unconditionally. The target address is the only operand of this instruction apart from the predicates. The jump is unconditional if the instruction is used without any predicates. However, \texttt{BRA} can be guarded by up to two predicates, making the branch conditional. For example, only threads whose \texttt{P0} is \texttt{false} and \texttt{P1} is \texttt{true} will jump to the target address when executing \texttt{@!P0 BRA P1, target}.

\subsection{CALL and RET Instructions}
NVIDIA uses registers instead of a stack-based mechanism to store the return address when calling a function. In this scheme, the compiler stores the return address in registers before executing the \texttt{CALL} instruction, typically by a \texttt{MOV} instruction. Within the function, the \texttt{RET} instruction uses the same registers to return to the caller function. \texttt{CALL} and \texttt{RET} instructions have modifiers that slightly change their behavior. 

\subsection{BMOV, BSSY, BSYNC, and BREAK Instructions}
Our study shows that the compiler inserts \texttt{BMOV}, \texttt{BSSY}, \texttt{BSYNC}, and \texttt{BREAK} instructions in the program flow to assist the control flow management mechanism in thread reconvergence after branches. These instructions are not visible to CUDA or PTX programmers and are only added to the program by the compiler.

A \texttt{BSYNC} instruction is used at reconvergence points to reunite threads of a warp after a branch divergence. Reconverging at \texttt{BSYNC} instead of IPDom points allows for reconvergence earlier or later than the IPDom point. Reconverging earlier can enhance the performance of some programs with unstructured control flow (details in Section \ref{sec:break_usage}) while reconverging later is necessary to avoid SIMT-induced deadlocks appearing in pre-Volta GPUs (explained in more detail in Section \ref{sec:deadlock_free_spinlock}). NVIDIA’s compiler analyzes the source code and inserts \texttt{BSYNC} at an appropriate point in the program flow. This point could be at the IPDom or other places.

A \texttt{BSYNC} instruction takes only one operand, a special-purpose register denoted by \texttt{B\textsubscript{x}}. The value stored in a \texttt{B\textsubscript{x}} register is crucial for understanding \texttt{BSYNC}. Unfortunately, the NVIDIA binary instrumentation tool, NVBit \cite{NVBit}, does not capture the value of these registers unlike general-purpose registers. However, \texttt{BMOV} instructions transfer values between \texttt{B\textsubscript{x}} and \texttt{R\textsubscript{x}} registers. Hence, By reading \texttt{R\textsubscript{x}} values we can get access to \texttt{B\textsubscript{x}} values. We realized that a \texttt{B\textsubscript{x}} register contains a mask. Further investigation revealed that this mask indicates which threads of a warp the \texttt{BSYNC} must reconverge. We call this mask a \textbf{\textit{reconvergence mask}} throughout this paper. For example, if reconvergence mask in \texttt{B0} is \texttt{1100}, instruction \texttt{BSYNC B0} would only reconverge threads 2 and 3.

Reading \texttt{B\textsubscript{x}} values also helped us grasp the semantics of \texttt{BSSY} instruction. A \texttt{BSSY} instruction initializes a \texttt{B\textsubscript{x}} register and specifies the reconvergence point via an instruction PC. This PC always points to a \texttt{BSYNC} instruction. \texttt{B\textsubscript{x}} is initialized with a mask showing which threads of a warp are active when the the \texttt{BSSY} instruction is executed. This mask represents the reconvergence mask since a branch that causes divergence is always preceded by a \texttt{BSSY} instruction. Hence, all threads of a warp that are indicated in the reconvergence mask have previously executed the \texttt{BSSY} instruction before diverging because of the branch. For example, if threads 2 and 3 of a warp execute \texttt{BSSY B0, 1000} before a branch, we must initialize \texttt{B0} to \texttt{1100}. A \texttt{BSYNC B0} at address 1000 reads \texttt{B0} and reunites threads 2 and 3.

We also discovered that the \texttt{BREAK} instruction is necessary to avoid deadlocks when the reconvergence point is not at IPDom points. By definition, all diverging threads must pass through the IPDom point before the program finishes. However, there is no guarantee that all diverging threads will pass through some of these reconvergence points that are not IPDom points. Therefore, some diverging threads in a warp may never reach their specified reconvergence point. Waiting for these threads at the reconvergence point creates a deadlock unless a \texttt{BREAK} instruction removes them from the reconvergence mask. Section \ref{sec:break_usage} elaborates on using BREAK instruction to avoid deadlocks when the reconvergence point is earlier than the IPDom point.

A \texttt{BREAK} instruction takes one or two predicates and a \texttt{B\textsubscript{x}} register. The predicates determine which threads of a warp must be removed from the reconvergence mask in the \texttt{B\textsubscript{x}} register. The removed threads will not be reunited with the other active threads that are present in the mask. For example, \texttt{@P0 BREAK !P1, B0} removes threads with a true \texttt{P0} and a false \texttt{P1} from the reconvergence mask in \texttt{B0}. After this, a \texttt{BSYNC B0} instruction will not wait for the removed threads unless some other instructions change \texttt{B0}.

\subsection{WARPSYNC Instruction}

A \texttt{WARPSYNC} instruction synchronizes threads of a warp, similar to \texttt{BSYNC} instructions. This instruction takes only one operand, which is a mask indicating which threads of a warp must be synchronized. This mask has the same definition as the \textbf{\textit{reconvergence mask}} in \texttt{BSYNC} instructions, so we refer to them with the same name. The reconvergence mask in \texttt{WARPSYNC} is stored in an \texttt{R\textsubscript{x}} register or an immediate value. For example, \texttt{WARPSYNC 1100} synchronizes threads 2 and 3 and is exactly the same as \texttt{WARPSYNC R0} when \texttt{R0} contains \texttt{1100} for both threads 2 and 3.

% Although both \texttt{WARPSYNC} and \texttt{BSYNC} instructions synchronize threads within the warp, they have fundamental differences:
% \begin{itemize}
%     \item The programmer inserts \texttt{WARPSYNC} instructions in the program flow, while the compiler inserts \texttt{BSYNC} instructions.
%     \item The programmer sets the reconvergence mask for \texttt{WARPSYNC} at programming time, but the reconvergence mask for \texttt{BSYNC} is determined at runtime.
%     \item \texttt{BSSY} instructions inform the control flow management mechanism about \texttt{BSYNC} instructions before their execution. However, there is no such notification for \texttt{WARPSYNC} instructions.
%     \item If we ignore all \texttt{BSYNC} instructions, the program will still be correct because they are only visible to the compiler, and program logic does not rely on them. However, \texttt{WARPSYNC} is inserted by the programmer and is part of the program logic, which cannot be ignored.
% \end{itemize}
\subsection{YILED Instruction}

We define the semantics of the \texttt{YIELD} instruction based on several key observations. We observed that threads within a warp only diverge at branches and reconverge at \texttt{BSYNC} or \texttt{WARPSYNC} instructions. For all other instructions, threads within a warp execute in a lockstep manner. Additionally, we observed that interleaved execution of different paths only happens as a consequence of execution a \texttt{YIELD} instruction.

Based on these observations, we concluded that the two paths of a branch are scheduled sequentially one after the other, unless a \texttt{YIELD} instruction is executed. This instruction causes the control flow to switch to a new path. A \texttt{YIELD} instruction does not have any operands to indicate which path must be scheduled next. We assume it must be the sibling path since it matches the hardware's control flow trace observed in our experiments and it does not require a costly microarchitecture. If no sibling path exists, execution continues with the instruction following \texttt{YIELD}.

This definition of \texttt{YIELD} relaxes the first constraint that pre-Volta GPUs have when scheduling diverging threads of a warp (see section \ref{sec:pre_volta_deadlock}). Based on this constraint, pre-Volta GPUs never interleave the execution of different paths, but post-Volta GPUs like Turing use \texttt{YIELD} instructions for that purpose. Therefore, \texttt{YIELD} can resolve some of the SIMT-induced deadlocks in pre-Volta GPUs caused by the first constraint.

We designed an experiment to verify that NVIDIA uses \texttt{YIELD} to solve some of these deadlocks. We removed \texttt{YIELD} from the binary of the source code shown in Figure 3. This source code is a well-known example of SIMT-induced deadlocks for pre-Volta GPUs (see section \ref{sec:pre_volta_deadlock}). The original program finishes on Turing GPUs, but after removing \texttt{YIELD}, it never finishes. This experiment verifies that Turing relies on the \texttt{YIELD} instruction to avoid certain SIMT-induced deadlocks present in pre-Volta GPUs. Since \texttt{YIELD} is only visible to the compiler, NVIDIA's compiler algorithms are crucial for solving these deadlocks by inserting the \texttt{YIELD} instruction at an appropriate point (more details in section \ref{sec:deadlock_free_spinlock}).

\section{Practical Applications of Turing Control Flow Instructions}
The control-flow management mechanism in Turing uses the control-flow instructions in various scenarios. This section explains their use in three practical cases: 1) thread reconvergence after nested branches. 2) reconverging earlier than the IPDom point, and 3) Spinlock implementation.

\subsection{Reconvergence after Nested Branches}

Nested branches may cause nested thread divergence, requiring multiple \texttt{B\textsubscript{x}} registers to store reconvergence masks. Theoretically, up to 31 \texttt{B\textsubscript{x}} registers may be needed, since when the 32 threads of a warp can diverge into 32 distinct paths. However, NVIDIA uses fewer \texttt{B\textsubscript{x}} registers and spills their values to \texttt{R\textsubscript{x}} registers to save hardware resources. \texttt{BMOV} instructions transfer values between \texttt{B\textsubscript{x}} and \texttt{R\textsubscript{x}} registers for this purpose. Research has shown that many \texttt{R\textsubscript{x}} registers are dead during runtime \cite{RF_Virtualization}, making them good candidates to temporarily store \texttt{B\textsubscript{x}} registers before their use at a reconvergence point.

\begin{figure}[t]
    \centering
    \begin{subfigure}[b]{0.34\columnwidth}
        \centering
        \includegraphics[width=\linewidth]{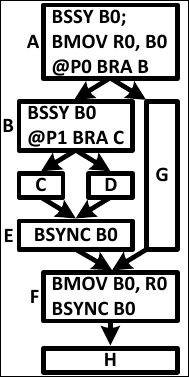}
        \caption{Control flow graph}
        \label{fig:nested_cfg}
    \end{subfigure}
    \hfill
    \begin{subfigure}[b]{0.60\columnwidth}
        \centering
        \includegraphics[width=\linewidth]{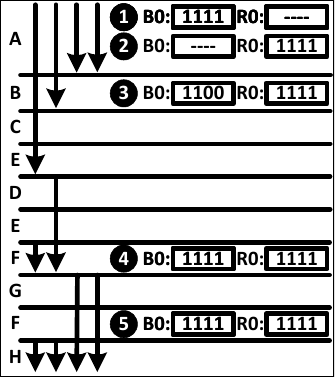}
        \caption{Execution}
        \label{fig:nested_execution}
    \end{subfigure}
    \caption{Sample of reconvergence after nested branches}
    \label{fig:nested_reconvergence}
\end{figure}

The order of using \texttt{BMOV}, \texttt{BSSY}, and \texttt{BSYNC} instructions is crucial in a nested thread divergence scenario. \texttt{BSSY} initializes a \texttt{B\textsubscript{x}} register, and \texttt{BSYNC} uses it to enforce reconvergence. The value in \texttt{B\textsubscript{x}} must be moved to an \texttt{R\textsubscript{x}} register after initialization and returned to \texttt{B\textsubscript{x}} before the reconvergence point. This data transfer is necessary if \texttt{B\textsubscript{x}} is going to be allocated to a new reconvergence mask that will be used to reunite the diverged threads after a nested branch.

Figure \ref{fig:nested_reconvergence} shows \texttt{BSSY}, \texttt{BSYNC}, and \texttt{BMOV} instructions in a sample program with two nested thread divergences. The control flow graph of this program is depicted in Figure \ref{fig:nested_cfg}, and how a sample warp updates registers during the execution is represented in Figure \ref{fig:nested_execution}. Basic blocks in the control flow graph may have many more instructions, but we only show the control flow instructions and their correct order within basic blocks. By means of multiple experiments, we observed that all active threads executing \texttt{BMOV} instructions read or write the same value in \texttt{R\textsubscript{x}} registers. Therefore, in this figure, we only represent the \texttt{R\textsubscript{x}} value for one of the active threads executing the \texttt{BMOV} instruction. We show a dead value in a register by a dashed line. In this example, reconvergence is enforced at \texttt{BSYNC} instructions, and the taken path is prioritized after branch divergence. We observed that, NVIDIA prioritizes the path most threads follow, but this is just an optimization. NVIDIA’s compiler generates a binary whose correct execution does not depend on which path, taken or not-taken, will be prioritized in runtime by the control flow management mechanism.

In this example, B0 is used in two \texttt{BSYNC} instructions: one in E that reconverges Threads 2 and 3, and another in F that reunites all threads. Each \texttt{BSYNC} instruction requires a \texttt{BSSY} instruction to initialize the B0 register with the reconvergence mask. The \texttt{BSYNC} in F needs a \texttt{BSSY} in A that initializes B0 with 1111 (\circled{1}), while the \texttt{BSYNC} in E requires a \texttt{BSSY} in B that initializes B0 with 1100 (\circled{3}). Since B is executed after A, it overwrites the value in B0. This value, the reconvergence mask, is needed later in F to reconverge all threads. To avoid losing this value when B is executed, a \texttt{BMOV} in A copies B0 to R0 (\circled{2}). The value remains in R0 until the \texttt{BSYNC} in F requires it. Before that, a \texttt{BMOV} in F retrieves this value from R0 and writes it back to B0 (\circled{4}, \circled{5}). This example illustrates how \texttt{BMOV} uses the R0 register as a backup for the B0 register in this nested thread divergence scenario. By doing this, NVIDIA avoids increasing the number of \texttt{B\textsubscript{x}} registers needed to handle the worst possible nesting scenario. NVIDIA's compiler inserts \texttt{BMOV} instructions in the program to spill \texttt{B\textsubscript{x}} values to \texttt{R\textsubscript{x}} registers if there are not enough \texttt{B\textsubscript{x}} registers available.

\subsection{Reconvergence Earlier than IPDom} \label{sec:break_usage}

\begin{figure}[t]
    \centering
    \begin{subfigure}[b]{0.48\columnwidth}
        \centering
        \includegraphics[width=\linewidth]{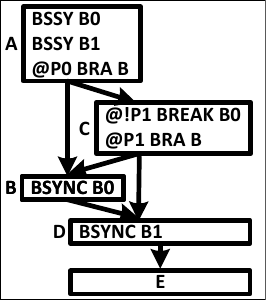}
        \caption{Control flow graph}
        \label{fig:early_cfg}
    \end{subfigure}
    \hfill
    \begin{subfigure}[b]{0.48\columnwidth}
        \centering
        \includegraphics[width=\linewidth]{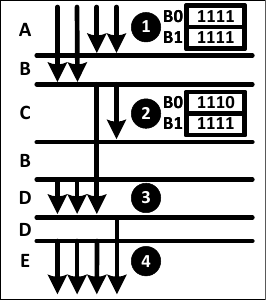}
        \caption{Runtime register updates}
        \label{fig:early_exec}
    \end{subfigure}
    \caption{Sample of earlier reconvergence than IPDom}
    \label{fig:early_reconvergence}
\end{figure}

Reconvergence using a \texttt{BSYNC} instruction has the advantage of allowing for reconvergence earlier or later than the IPDom point. Reconverging earlier can enhance the performance of some programs with unstructured control flow. However, in scenarios where reconvergence occurs earlier than the IPDom, incorrect use of control-flow instructions can lead to a deadlock. This deadlock may occur because an early reconvergence point is not an IPDom point where all diverged threads must pass through before finishing. In other words, some diverged threads may never reach an early reconvergence point, hence, waiting for them at the reconvergence point can cause a deadlock.

Figure \ref{fig:early_reconvergence} illustrates a sample program where early reconvergence is possible. Figure \ref{fig:early_cfg} depicts the program's control flow graph, while Figure \ref{fig:early_exec} shows register updates during the execution of a single warp. Although each basic block may contain many more instructions, we display only the control-flow instructions for simplicity. NVIDIA's compiler inserts the control-flow instructions within each basic block in the same order as shown in the figure. We assume the taken path is executed before the not-taken path, and threads within a warp are reconverged after \texttt{BSYNC} instructions.

In this example, \texttt{D} is the IPDom point for \texttt{A} because this is the earliest point that all threads in the warp pass through it before finishing. However, thread 0 never passes through \texttt{B} after the divergence in \texttt{A} and executes \texttt{C}, \texttt{D}, and \texttt{E} before finishing. Therefore, \texttt{B} cannot be an IPDom point for \texttt{A}, as all diverged threads must pass through the IPDom point by definition. However, \texttt{B} serves as an early reconvergence point for threads 1, 2, and 3 of the warp since all of them can be reunited at this point before executing D.

This example contains two \texttt{BSYNC} instructions to reconverge threads within a warp that diverge after A: one is inserted in \texttt{B} and the other in \texttt{D}. These \texttt{BSYNC} instructions read their reconvergence masks from \texttt{B0} and \texttt{B1}, which are initialized in \texttt{A} before the divergence. Two \texttt{BSSY} instructions in \texttt{A} initialize \texttt{B0} and \texttt{B1} to \texttt{1111} since all threads of the warp execute the BSSY instructions together (\circled{1}). If the value in \texttt{B0} does not change, this program will never finish due to a deadlock. The deadlock happens because \texttt{B0} contains \texttt{1111} and is used by \texttt{BSYNC} in \texttt{B}, which means all threads of the warp must reconverge after \texttt{B}. However, thread 0 in the warp never executes \texttt{B}. Therefore, all other threads of the warp waiting for thread 0 to execute \texttt{B} would be blocked forever.

To solve this problem, a \texttt{BREAK} instruction is inserted in \texttt{C}. The \texttt{BREAK} removes all threads in \texttt{C} whose \texttt{P1} is false from the reconvergence mask in \texttt{B0}. These threads will diverge to \texttt{D} after executing the branch instruction in \texttt{C} and will never execute \texttt{B} again. In this example, we assumed only thread 0 has a false \texttt{P1} when executing the \texttt{BREAK} in \texttt{C}. Removing thread 0 from \texttt{B0} changes its value to \texttt{1110} in \texttt{B0}, which is the reconvergence mask in \texttt{B} (\circled{2}). Therefore, only threads 1, 2, and 3 are reunited after \texttt{B} (\circled{3}), and they never wait for thread 0. Thread 0 joins them after \texttt{D} (\circled{4}), where the program does not have any divergence afterward. This example shows that if we add \texttt{BREAK} in \texttt{C}, early reconvergence is enforced in \texttt{B}, and the program finishes correctly without any deadlock. This example emphasizes that the compiler can assist the control flow management mechanism in reuniting threads earlier than the IPDom points by simply inserting the control-flow instructions in the right locations in the program.

\subsection{Spinlock Implementation}
\label{sec:deadlock_free_spinlock}

\begin{figure}[t]
    \centering
    \begin{subfigure}[b]{0.58\columnwidth}
        \centering
        \includegraphics[width=\linewidth]{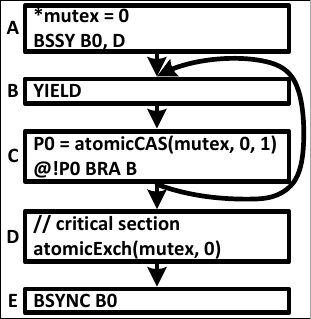}
        \caption{Control flow graph}
        \label{fig:spinlock_CFG}
    \end{subfigure}
    \hfill
    \begin{subfigure}[b]{0.40\columnwidth}
        \centering
        \includegraphics[width=\linewidth]{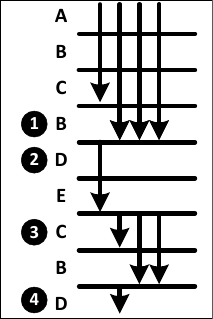}
        \caption{Sample warp execution}
        \label{fig:spinlock_EXEC}
    \end{subfigure}
    \caption{Sample of spinlock implementation in Turing}
    \label{fig:turing_spinloc}
\end{figure}

A CUDA spinlock implementation, as shown in Figure \ref{fig:pre_volta_deadlock}, causes a well-known SIMT-induced deadlock in pre-Volta GPUs due to two constraints: 1) pre-Volta GPUs enforce reconvergence at the IPDom points, and 2) they execute diverged paths serially, one after another (more details in section \ref{sec:pre_volta_deadlock}). Turing prevents this deadlock by removing these constraints with the compiler's help. The compiler does not place \texttt{BSYNC} at the IPDom point if it causes a deadlock and also inserts \texttt{YIELD} to switch to a sibling path when executing the same path would lead to a deadlock.

Figure \ref{fig:turing_spinloc} illustrates the crucial role of \texttt{BSYNC} and \texttt{YIELD} in preventing deadlock in the spinlock implementation depicted in Figure \ref{fig:pre_volta_deadlock}. Figure \ref{fig:spinlock_CFG} shows the control-flow graph of this implementation augmented with Turing's control-flow instructions. Figure \ref{fig:spinlock_EXEC} illustrates a possible  execution by a sample warp, assuming the taken path is executed before the not-taken path after divergence. Before the taken path finishes, \texttt{YIELD} intervenes and switches execution to the not-taken path. The \texttt{YIELD} is necessary to avoid deadlock we observed that if we remove it from the binary of the program, its execution never finishes.

This program contains a critical section in D, right after a loop. Within the loop, threads compete to acquire a lock and get permission to execute the critical section. Only one of them acquires the lock and sets \texttt{P0} to \texttt{true} when executing C. This thread diverges from the other threads in its warp and exits the loop to execute the critical section. After the critical section, it releases the lock and allows another thread to acquire the lock and enter the critical section.

In this example, thread 3 acquires the lock and diverges from the other threads in the warp but does not execute the critical section immediately because the taken path has higher priority and must be executed first. Hence, threads 0, 1, and 2 jump to the beginning of the loop and execute the \texttt{YIELD} (\circled{1}). If this \texttt{YIELD} did not exist, these threads could never acquire the lock because thread 3 still holds it. Therefore, these threads would be trapped in the loop forever. However, the \texttt{YIELD} solves this issue and switches the execution to D (\circled{2}), the sibling path. Thread 3 executes the critical section in D and releases the lock. Then, it executes the \texttt{BSYNC} in E, which is used to reconverge all threads. After that, thread 3 remains blocked until all other threads execute this \texttt{BSYNC}. In this scenario, all other threads continue execution from C (\circled{3}), the next instruction after \texttt{YIELD}. This time, one of them can acquire the lock because thread 3 has already released it. In this example, thread 2 acquires the lock and executes the critical section (\circled{4}).

If the \texttt{BSYNC} in E was placed before releasing the lock, it would also cause a deadlock. The deadlock would happen because the thread that has the lock waits infinitely to reconverge with other threads in the warp, but the other threads are stuck in the loop because none of them can acquire the lock and exit the loop. In this program, the IPDom point is right after the branch, which is before releasing the lock. Therefore, after detecting the deadlock scenario, NVIDIA's compiler decides to put \texttt{BSYNC} later than the IPDom point in E after the lock was released.
\section{Hanoi Microarchitecture}

\begin{figure}[t]
    \centering
    \includegraphics[width=\linewidth]{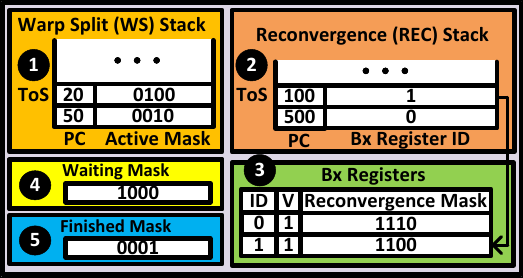}
    \caption{Hanoi Microarchitecture}
    \label{fig:hanoi_microarchitecture}
\end{figure}

Figure \ref{fig:hanoi_microarchitecture} illustrates our proposed design called Hanoi for a control flow management mechanism in Turing. This design comprises two stacks: the Warp Split (WS) (\circled{1}) and Reconvergence (REC) (\circled{2}) stacks. It also includes a few \texttt{B\textsubscript{x}} registers (\circled{3}) and two masks: the waiting (\circled{4}) and finished (\circled{5}) masks.

The WS stack has one entry per path to be executed, each containing a PC and an active mask. The paths in the WS stack are executed in the stack order, and the entry at the top of the stack corresponds to the path currently in execution. The active mask indicates which threads in the warp are following this path, while the PC specifies the next instruction these threads must execute. In this figure, thread 2 in the warp is set to execute instruction 20 in one path, and thread 1 is set to execute instruction 50 in another path.

The REC stack has one entry per reconvergence point, each containing a PC and a \texttt{B\textsubscript{x}} register ID. The reconvergences occur with their order in the REC stack. The entry at the top of the stack corresponds to the path currently in execution. The PC points to the instruction that threads must execute after reconvergence, and the \texttt{B\textsubscript{x}} register ID refers to the \texttt{B\textsubscript{x}} register containing the reconvergence mask. The reconvergence mask indicates which threads in the warp must reunite at this reconvergence point. A \texttt{B\textsubscript{x}} register only contains a valid reconvergence mask when its valid (V) bit is set to 1. This figure shows two reconvergence points: one at PC 100 and the other at PC 500. The entry in the REC stack for the reconvergence point at PC 100 refers to \texttt{B\textsubscript{1}}, which contains the reconvergence mask indicating that threads 2 and 3 must reconverge at this point. The entry for the reconvergence at PC 500 refers to \texttt{B\textsubscript{0}}, containing a reconvergence mask indicating that threads 2 and 3 must join thread 1 at this point.

The waiting mask indicates which threads are waiting at the current reconvergence point (i.e., the one at the top of the REC stack). In this figure, thread 3 has already reached its reconvergence point at PC 100. The finished mask tracks which threads have executed the \texttt{EXIT} instruction and are already finished. In this figure, thread 0 has already finished.

\subsection{Managing \texttt{B\textsubscript{x}} Registers}

A \texttt{BSSY} instruction initializes \texttt{B\textsubscript{x}} registers. It uses the \texttt{B\textsubscript{x}} register ID in its operands to address a specific \texttt{B\textsubscript{x}} register. The \texttt{B\textsubscript{x}} register is initialized with the active threads in the warp executing the \texttt{BSSY} instruction. These threads are exactly the ones indicated in the active mask of the top entry in the WS stack. Hence, when \texttt{BSSY} is executed, this active mask is copied to the specified \texttt{B\textsubscript{x}} register, and its valid bit is set to 1.

A \texttt{BREAK} instruction updates a \texttt{B\textsubscript{x}} register. It addresses a \texttt{B\textsubscript{x}} register by its ID and removes specific threads from the reconvergence mask stored within it. This update is possible because reconvergence points in the REC stack read their reconvergence masks from \texttt{B\textsubscript{x}} registers. In other words, they indirectly refer to a \texttt{B\textsubscript{x}} register containing their reconvergence mask. If the reconvergence mask were stored directly in the REC entries, removing threads from the reconvergence mask of an entry not at the top of the stack would be impossible.

A \texttt{BMOV} instruction transfers \texttt{B\textsubscript{x}} registers to \texttt{R\textsubscript{x}} registers or vice versa. When a \texttt{BMOV} instruction moves a \texttt{B\textsubscript{x}} register to an \texttt{R\textsubscript{x}} register, it invalidates the \texttt{B\textsubscript{x}} register. Upon returning the value to the \texttt{B\textsubscript{x}} register, its valid bit is set to 1 again. \texttt{BMOV} instructions are used for sharing \texttt{B\textsubscript{x}} registers among different reconvergence points. In other words, multiple entries in the REC stack may refer to the same \texttt{B\textsubscript{x}} register. However, an entry only reads a \texttt{B\textsubscript{x}} register when it is at the top of the REC stack. Therefore, a \texttt{BMOV} instruction can move a value from a \texttt{B\textsubscript{x}} register to an \texttt{R\textsubscript{x}} register, provided the value is moved back before the reconvergence point which needs this value becomes the top entry in the REC stack and this reconvergence point is reached in the program.

An \texttt{EXIT} instruction completes the execution for threads. Once an \texttt{EXIT} instruction is executed, Hanoi removes finished threads from all \texttt{B\textsubscript{x}} registers and adds them to the finished mask. Finished threads in this mask must be removed from any reconvergence mask read from \texttt{R\textsubscript{x}} registers before it is written to \texttt{B\textsubscript{x}} registers. This step is necessary to ensure correct control flow, as some threads might have already finished while a reconvergence mask was stored in the \texttt{R\textsubscript{x}}.

When a reconvergence occurs, the reconvergence mask for the reconvergence point, which is in a \texttt{B\textsubscript{x}} register, is no longer needed. Therefore, the \texttt{B\textsubscript{x}} register containing the reconvergence mask is invalidated.

\subsection{Managing REC Stack}

When reconvergence occurs, Hanoi pops the top entry from the REC stack, retrieves the reconvergence mask from its \texttt{B\textsubscript{x}} register, and pushes a new entry into the WS stack. This new entry contains the \texttt{PC} and reconvergence mask from the top entry of the REC stack. Thus, execution continues from the instruction after the reconvergence point for all threads in the reconvergence mask.

Hanoi uses the waiting mask to determine when it must reconverge threads, which is crucial for the correct traversal of the control flow graph during execution. Before scheduling the top entry of the WS stack, Hanoi checks if the reconvergence mask of the top entry in the REC stack is valid and fully appears in the waiting mask. If it does, all threads in the reconvergence mask have reached the reconvergence point, and it is time to reconverge them. Hanoi never reconverges threads if the \texttt{B\textsubscript{x}} register of the top entry in the REC stack is invalid since this Bx register might have been overwritten and used by another reconvergence point.

Hanoi uses the REC stack for reconvergence at \texttt{BSYNC} or \texttt{WARPSYNC} instructions. While both instructions reconverge threads within a warp, they have significant differences that require different handling.

Before each \texttt{BSYNC} instruction, there is a corresponding \texttt{BSSY} instruction that specifies the reconvergence \texttt{PC} and the ID of the \texttt{B\textsubscript{x}} register holding the reconvergence mask. After executing \texttt{BSSY}, Hanoi pushes an entry onto the REC stack with the specified reconvergence \texttt{PC} and \texttt{B\textsubscript{x}} register ID.

In contrast, \texttt{WARPSYNC} has no prior instruction like \texttt{BSSY} and lacks a \texttt{B\textsubscript{x}} operand. For \texttt{WARPSYNC}, Hanoi must allocate an empty \texttt{B\textsubscript{x}} register and initialize it with the reconvergence mask. We assume an empty \texttt{B\textsubscript{x}} register is always available; if not, the compiler can spill them to \texttt{R\textsubscript{x}} registers using \texttt{BMOV} instructions. The \texttt{PC} of the instruction following \texttt{WARPSYNC} becomes the reconvergence \texttt{PC} in its REC stack entry, as this is where threads continue execution after reconvergence. The reconvergence \texttt{PC} and the ID of the allocated \texttt{B\textsubscript{x}} register form the entry pushed to the REC stack for a \texttt{WARPSYNC} instruction. However, this is done only for the first subset of threads in the reconvergence mask executing the \texttt{WARPSYNC} instruction. This entry is already on the REC stack for other subsets, so Hanoi does not push it again.

Having a separate stack for tracking reconvergence points, the REC stack, allows for efficient handling of \texttt{WARPSYNC} instructions in Hanoi, as described above. Supporting \texttt{WARPSYNC} instructions is either impossible or highly inefficient in alternative mechanisms that use only one stack. This is because, in mechanisms using a single stack, it is crucial to specify the reconvergence point before divergence happens to update the stack correctly and reconverge threads at that point. However, when a \texttt{WARPSYNC} is executed, we only know that threads might have diverged somewhere before \texttt{WARPSYNC}, without knowing exactly where. Hence, it is not possible or inefficient to update the stack so that reconvergence occurs correctly when the \texttt{WARPSYNC} is executed.

\subsection{Managing WS Stack}

Hanoi may push or pop entries onto the WS stack due to \texttt{BSYNC}, \texttt{WARPSYNC}, \texttt{EXIT}, \texttt{BRA}, and \texttt{YIELD} instructions. For other instructions, only the PC for the top entry in the WS stack needs to be updated. Some instructions, such as \texttt{CALL} and \texttt{RET}, change the PC to jump to or return from a function. However, incrementing the PC to point to the following instruction is sufficient for most instructions. 

A \texttt{BRA} instruction may cause thread divergence, resulting in two entries being pushed: one for the taken path and another for the not-taken path. Our observations indicate that the path followed by the majority of threads is executed first, so Hanoi pushes this path after the other one.

An entry must be popped when a path finishes. A path is considered finished when \texttt{BSYNC}, \texttt{WARPSYNC}, or \texttt{EXIT} instructions are executed. After executing \texttt{BSYNC} or \texttt{WARPSYNC}, Hanoi pops the entry and adds its active threads to the waiting mask. However, for EXIT instruction, an entry is popped from the WS stack only when all threads in the path execute the \texttt{EXIT}. Some threads may be predicated off and do not execute \texttt{EXIT}. For them, the instruction following the \texttt{EXIT} will be executed in the next cycle. Hanoi adds threads executing \texttt{EXIT} to the finished mask.

Hanoi switches to a sibling path when a \texttt{YIELD} instruction is executed, provided a sibling exists. Switching to a non-sibling path generates a wrong control flow. If a sibling path exists, its entry is immediately below the top entry in the WS stack. Therefore, upon executing a \texttt{YIELD} instruction, Hanoi simply swaps the top two entries in the WS stack to switch to the sibling path. However, a path may have an entry below it that does not belong to its sibling. Paths are siblings only if they share a common reconvergence point, which is always represented by the top entry in the REC stack. Hence, to determine if the top two paths of the WS stack are siblings, Hanoi compares the union of their active mask with the reconvergence mask of the entry at the top of the REC stack. If this union is a subset of the  reconvergence mask, the paths are siblings; otherwise, they are not. For example, in Figure \ref{fig:hanoi_microarchitecture}, threads 1 and 2 are in the top two entries in the WS stack. However, these two paths are not siblings because the reconvergence mask for the top entry in the REC stack does not include thread 1. Therefore, executing a \texttt{YIELD} instruction does not switch to any other path and behaves like a NOP.

\section{Methodology}

\begin{table}[t]
    \centering
    \begin{tabularx}{\columnwidth}{|X|}
        \hline
        \textbf{Rodinia} \\
        \hline
        B+tree (BTR), Backprop (BKPR), BFS (RBFS), dwt2d (DWT), Gaussian (GAUS), Hotspot (HOTS), Kmeans (KMNS), LUD, NN, NW, Particlefilter\_Float (PARF), Particlefilter\_Naive (PARN), Pathfinder (PFN), SRAD, lavaMD (LAVA) \\
        \hline
        \textbf{ISPASS} \\
        \hline
        AES, BFS, LIB, LPS, NN, NQU, STO, WP \\
        \hline
        \textbf{Lonestar} \\
        \hline
        BFS (LBFS), BFS-Atomic (BFSA), BFS-WLC (BFSW), DMR, NSP, SSSP \\
        \hline
        \textbf{GraphBIG} \\
        \hline
        Betweenness\_Centr (BETC), BFS\_Topo\_Thread\_Centric (BFST), BFS\_Data\_Warp\_Centric (BFSD), BFS\_Topo\_Unroll (BFSU), Connected\_Comp (CC), Degree\_Centr (DC), Kcore (KCOR), SSSP\_Topo\_Thread\_Centric (SSTT), Triangle\_Count (TC) \\
        \hline
        \textbf{Tango} \\
        \hline
        AlexNet (AN), GRU, LSTM \\
        \hline
    \end{tabularx}
    \caption{List of beenchmarks studied in this work}
    \label{tab:benchmarks}
\end{table}

\begin{table}[tb]
  \centering
  \small
  \begin{tabular}{|l|l|}
    \hline
    \textbf{\#SMs} & 30\\
    \hline
    \textbf{\#Threads/Warps per SM } & 1024 / 32\\
    \hline
    \textbf{\#sub-cores per SM } & 4\\
    \hline
    \textbf{RF size per SM} & 256KB\\
    \hline
    \textbf{\#Issue Schedulers per SM} & 4\\
    \hline
    \textbf{Issue Scheduling Policy} & GTO\\
    \hline
    \textbf{L2 size} & 3MB\\
    \hline
    \textbf{L1/Shared Memory per SM} & 64KB\\
    \hline
  \end{tabular}
  \caption{Baseline GPU configuration used in this work}
  \label{tab:configs}
\end{table}

% \begin{figure*}[!t]
%     \centering
%     \includegraphics[width=\textwidth]{figs/CFI_SIMD.png}
%     \caption{Percentage of control flow instructions in the benchmarks and their SIMD efficiency when running on our Turing GPU}
%     \label{fig:CFI_SIMD}
% \end{figure*}

\begin{figure*}[t]
    \centering
    \begin{minipage}[t]{0.48\textwidth}
        \centering
        \includegraphics[width=\linewidth]{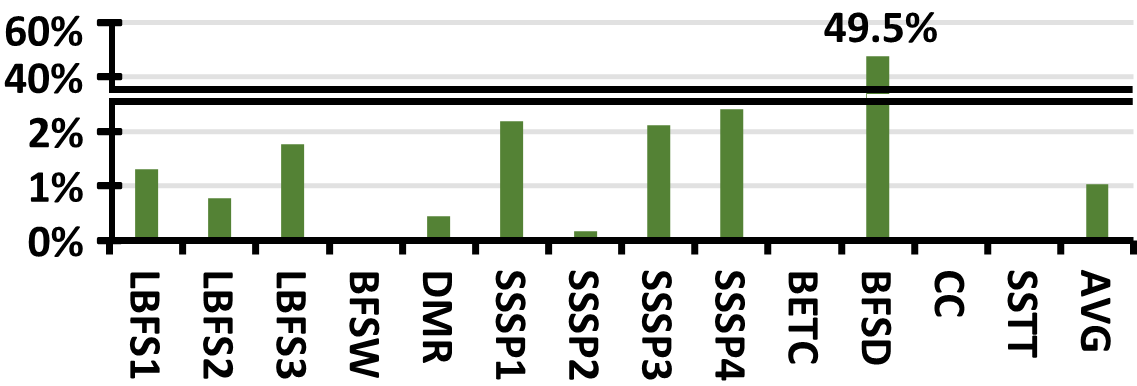}
        \caption{Percentage discrepancy between the control flow traces of Hanoi and Turing. For the other 43 benchmarks not shown in the figure the discrepancy is zero.}
        \label{fig:Levenshtein}
    \end{minipage}%
    \hfill
    \begin{minipage}[t]{0.48\textwidth}
        \centering
        \includegraphics[width=\linewidth]{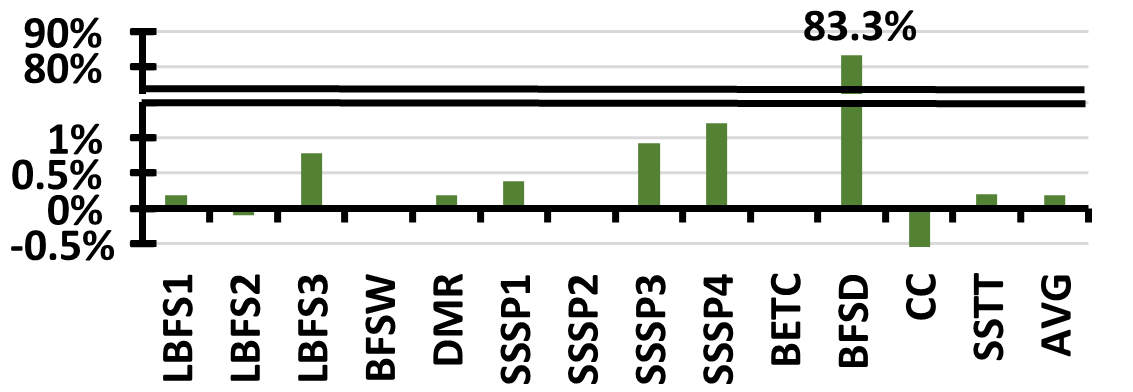}
        \caption{Percentage difference in IPC between Hanoi and Turing. For the other 43 benchmarks not shown in the figure the difference is zero.}
        \label{fig:IPC}
    \end{minipage}
\end{figure*}

We utilized NVIDIA tools to generate the native assembly source code (SASS), control flow graph (CFG), and traces for a large variety of well-known benchmarks, summarized in Table \ref{tab:benchmarks}. Specifically, we used cuobjdump \cite{TuringSASS} for SASS code generation, nvdisasm \cite{TuringSASS} for CFG generation, and NVBit \cite{NVBit} for trace generation. For some benchmarks, different input data sets were used, indicated by a number in front of their names throughout this paper. Through thorough analysis of these benchmarks, we identified patterns, scenarios, and use cases illustrating how control-flow instructions are defined, utilized, and handled by the control-flow management mechanism in Turing GPUs. This analysis allowed us to formulate a set of rational and intuitive hypotheses regarding the semantics of the control-flow instructions and detailed policies of the control flow management mechanism in Turing.

Subsequently, we developed a checker program to validate our hypotheses. We created parsers for the CFG, SASS code, and traces. After parsing these files, we loaded them into the checker, which verified whether any of our hypotheses were violated in the traces obtained from the actual hardware. For example, when our hypothesis is that interleaving happens only after a YIELD instruction and in the traces interleaving happens at some other point, it would be a violation. If a violation occurred, a log file was generated to provide insights into the cause of the violation. We then refined our hypotheses and repeated the process. Our objective was to develop a set of hypotheses that are never violated across all benchmarks, thereby explaining the semantics of the control flow instructions and the detailed policies employed in the Turing control flow management mechanism.

Given that NVIDIA uses runtime heuristics, it is extremely challenging to discover and incorporate all of them into our hypotheses set. However, we managed to uncover many of them, making our design, Hanoi, work very close to Turing's design. To gauge our proximity, we designed Hanoi based on our hypotheses and generated its control flow trace. We then compared this trace to Turing's trace and fine-tuned our design to minimize any discrepancies. To evaluate the impact of these discrepancies on performance, we used Hanoi's traces in a well-known trace-driven simulator called Accel-Sim \cite{Accel-Sim} and compared the measured performance with the baseline using Turing's traces. For our performance simulations, we utilized the configuration of the NVIDIA RTX 2060, as detailed in Table \ref{tab:configs}.

\section{Evaluation}

We developed a set of assumptions about the semantics of control-flow instructions and the detailed policies used in Turing's control flow management mechanism. Based on these assumptions, we designed Hanoi and refined both the assumptions and the design to closely match Turing's architecture, to end up with the designed described in section \ref{fig:hanoi_microarchitecture}.

To evaluate how close Hanoi is to Turing's design, we compared their control flow traces. A control flow trace shows the sequence of instructions each warp executes from the program's start to end. If a specific warp has an identical control flow trace in both Hanoi and Turing, it means that both designs executed exactly the same threads of the warp together at every cycle. However, significant differences in the sequences do not necessarily indicate substantial differences in the designs. A minor change such as prioritizing the taken path over the not-taken path after a branch divergence, can result in a significantly different control flow trace.

The control flow trace is essentially a sequence of instructions. To compare them, we used the Levenshtein distance \cite{levenshtein}, which measures the minimum number of insertions, deletions, or updates needed to transform one trace into the other. We divided this distance by the size of the trace to compute a metric showing the percentage discrepancy between the two traces.

We analyzed 59 program executions: 41 different benchmarks and 18 that were some of this benchmarks varying input data. Among these 59, 46 produced a 0\% discrepancy, indicating that Hanoi and Turing generated exactly the same control flow trace. The discrepancies for the remaining executions are depicted in Figure \ref{fig:Levenshtein}. Except for BFSD, all benchmarks exhibit negligible discrepancy percentages below 2.4\%, with an average discrepancy of just 1\%. The high discrepancy for BFSD (49.5\%) is due to a runtime heuristic in Turing that Hanoi does not support. Hanoi enforces reconvergence at all \texttt{BSYNC} instructions, while Turing in some rare occasions ignores these reconvergences for performance reasons.

Figure \ref{fig:IPC} shows the impact of these discrepancies on performance. We computed the relative IPC difference between Hanoi and Turing. On average, this difference is 0.2\%, which is negligible. Except for BFSD, all other benchmarks show a performance impact below 1.2\%. BFSD, however, exhibits an 83.3\% performance gain in Hanoi. This gain results from Hanoi's enforcement of reconvergence at all \texttt{BSYNC} instructions, which improved SIMD unit utilization by 31.9\%, leading to a significant performance improvement. 

\subsection{Hardware Overhead}

Hanoi is a lightweight scheme that is seamlessly integrated into the GPU's pipeline, as shown in Figure \ref{fig:SIMT_Core}. It does not require an increase in the I-Buffer size or significant overhead in the scoreboard. The scoreboard only needs to be extended to track the dependencies between a few \texttt{B\textsubscript{x}} registers. The WS stack in Hanoi requires a maximum of 32 entries to support the scenario where all 32 threads of a warp diverge. In this case, the REC stack needs 31 entries to reconverge these threads. Managing WS and REC stacks has also low-cost since they are managed as stacks. Each entry in the WS stack is just a PC and a 32-bit mask, making it simpler than the SIMT-Stack, which requires a reconvergence PC on top of these fields. While the SIMT-Stack compares the reconvergence PC with the PC to identify when to pop an entry, Hanoi pops an entry when \texttt{WARPSYNC}, \texttt{BSYNC}, or \texttt{EXIT} is executed. Each entry in the REC stack is also very cost-effective; it consists of just a PC and an index to a \texttt{B\textsubscript{x}} register. Since \texttt{B\textsubscript{x}} registers can be shared among REC entries, only a few of them are needed. For example, in a system with 8 \texttt{B\textsubscript{x}} registers, 3 bits for indexing is sufficient. The waiting and finished masks are also only 32-bit masks. Overall, the total storage required by Hanoi for a design with 8 \texttt{\textsubscript{x}} registers is 432 bytes, which is negligible and almost 43\% less storage than that needed for a SIMT-Stack.

\section{Related Work}
GPU vendors such as NVIDIA have never fully disclosed their control flow management mechanisms. However, they have revealed some aspects in early publications \cite{Tesla} and documents \cite{CudaGuide}. Researchers have discovered more details through microbenchmarking \cite{Dem_GPU_Micro}, leading to the widespread acceptance of a SIMT-Stack \cite{RPU, DWF} design as the baseline. Based on this, researchers developed performance simulators such as GPGPU-Sim \cite{GPGPU-Sim_3, GPGPU_SIM_Manual}, which execute PTX \cite{PTX} instructions, sufficient for early GPUs and traditional workloads. However, simulating modern workloads that use closed-source libraries like cuDNN \cite{cuDNN} and cuBLAS \cite{cuBLAS} in PTX is impossible. To address this, researchers developed trace-driven simulators such as Accel-Sim \cite{Accel-Sim} that support SASS \cite{TuringSASS} instructions.

Accel-Sim relies on traces for control flow and does not model the functionality of the control flow mechanisms due to insufficient details for modern GPUs like Volta \cite{Volta_Whitepaper}. Volta introduced a new feature called Independent Thread Scheduling \cite{independent_tsch, Volta_Whitepaper}, resulting in considerable changes to NVIDIA's execution model, suggesting substantial changes in control flow management mechanisms. Despite this, public information is unavailable, and researchers who have demystified modern GPUs such as Volta \cite{Dissecting_Volta} and Turing \cite{dissecting_Turing} have not disclosed details of this mechanism.

To the best of our knowledge, this work is the first to describe the semantics of the control-flow instructions encountered in modern NVIDIA GPU binaries and propose an implementation of a control-flow mechanism that supports them, Hanoi. Supporting these instructions required significant differences in Hanoi's microarchitecture compared to other alternatives in the literature.

For example, Hanoi executes one path and can switch to its sibling after a \texttt{YIELD} instruction. This software-controlled interleaving requires simple hardware and provides a more predictable control flow that can be leveraged to optimize other components in the GPU. Other proposals support interleaving, but they all employ fine-grained interleaving during runtime, requiring much more complex hardware. DWS \cite{DWS} needs a table and a path scheduler instead of the WS stack in Hanoi. Unlike Hanoi, it ignores some reconvergence points and attempts to reconverge warp splits having the same PC during runtime. It is sensitive to the path scheduling policy, and the opportunity for reconvergence might be missed. Dual-Path \cite{Dual_Path} addresses this issue by extending SIMT-Stack entries to store two active paths in each entry. At any cycle, any of the two paths on top of stacks can be scheduled. However, this scheme cannot support \texttt{BREAK} and \texttt{WARPSYNC} because \texttt{BREAK} needs to modify the reconvergence mask, which may not be at the top of the stack, and \texttt{WARPSYNC} lacks a prior instruction like \texttt{BSSY} to know when to push on top of the stack for reconvergence. Multi-Path (MP) \cite{Multi-Path} allows scheduling any path but requires I-Buffer slots and scoreboards per path, which is a huge cost avoided in Hanoi. Warp splits and reconvergence points must also be stored in a random access memory and content addressable memory instead of the two stacks used in Hanoi. Furthermore, it does not propose any mechanism to support \texttt{WARPSYNC}, \texttt{YIELD}, \texttt{BREAK} and \texttt{EXIT}. AWARE \cite{MIMD_Sync} avoids some costs of MP but still does not propose any mechanism for these unsupported instructions. Subwarp interleaving \cite{subwarp_interleaving} has a considerably different design than Hanoi and supports fine-grained interleaving. However, the authors state that the significant cost of this scheme did not justify its benefits for a commercial product. This is not the case for Hanoi; it is designed based on real hardware traces and is lightweight and cost-effective.

Hanoi is a much more lightweight scheme compared to other alternatives. It is the first design to store reconvergence masks in \texttt{B\textsubscript{x}} registers. These \texttt{B\textsubscript{x}} registers can be modified to support  \texttt{BREAK} instructions. They can be transferred to \texttt{R\textsubscript{x}} registers and shared among reconvergence points. The compiler manages the transfers, keeping the hardware simple. Sharing \texttt{B\textsubscript{x}} registers also reduces the cost of storing metadata needed for reconverging threads. Hanoi is the first design that does not require storing a reconvergence PC per path and per reconvergence point unlike all other alternative designs \cite{DWF,DWS,Dual_Path,RPU,TBL_Compaction,MIMD_Sync,Multi-Path,Large_Warp}. It also does not need to store pending masks per reconvergence point similar to MP or AWARE. Instead, it leverages compiler assistance and uses simpler hardware mechanisms to reconverge threads.

Other proposals that can schedule threads from different warps \cite{DWF, TBL_Compaction, Large_Warp} are completely different from Hanoi, which only schedules threads within a warp. Other proposed mechanisms for reconverging threads at points other than the IPDom points are mechanisms sensitive to path scheduling policies \cite{Thread_Frontiers,Multi-Path,DWS}, or they use profiling \cite{TBL_Compaction} or oracle information \cite{DWF}. Hanoi uses \texttt{BREAK} and other control flow instructions to guarantee earlier reconvergence when possible. We do not know how the compiler inserts these instructions, but this mechanism is highly similar to speculative reconvergence \cite{speculative_rec} in the literature. Delaying the reconvergences when there is a chance for deadlock has already been proposed \cite{MIMD_Sync}, but it is different in Hanoi. Unlike this previous proposal, Hanoi relies on the use of \texttt{YIELD} instructions to avoid deadlock. We have experimentally observed that this is how modern NVIDIA GPUs operate, although we do not know the detailed algorithms that NVIDIA's compiler uses to detect or insert these instructions.
\section{Conclusion}
In this work, we uncovered the semantics of control-flow instructions in the SASS ISA and detailed policies used in Turing's control-flow management mechanism. To achieve this goal, we used the information in the traces and binaries of commonly used benchmarks. Based on our findings and verified assumptions, we designed a control flow management mechanism for Turing called Hanoi. Hanoi is cost-effective and generates a control flow trace that is highly similar to Turing. The discrepancy percentage between the control flow traces of Hanoi and Turing is 1\% on average, which leads to less than 0.2\% relative IPC difference  for a large set of diverse benchmarks.
%%%%%%% -- PAPER CONTENT ENDS -- %%%%%%%%

%%%%%%%%% -- BIB STYLE AND FILE -- %%%%%%%%
\bibliographystyle{IEEEtranS}
\bibliography{refs}
%%%%%%%%%%%%%%%%%%%%%%%%%%%%%%%%%%%%

\end{document}